\documentclass[aps,twocolumn,showpacs,superscriptaddress,floatfix]{revtex4}

\usepackage[dvips]{graphicx}
\usepackage{xspace}
\usepackage{latexsym}
\usepackage{amssymb}

\newcommand{\pizero}  {\pi^{0}}
\newcommand{\nue}     {\nu_{e}}
\newcommand{\numu}    {\nu_{\mu}}
\newcommand{\nutau}   {\nu_{\tau}}
\newcommand{\nus}     {\nu_{s}}
\newcommand{\numubar} {\overline{\nu}_{\mu}}

\begin{document}

%%% Title and Authors %%%
\title{Measurement of single $\pizero$ production
in neutral current neutrino interactions with water
by a 1.3\,GeV wide band muon neutrino beam}
\date{\today}

\newcommand{\Osaka}{\affiliation{ Department of Physics, Osaka
    University, Toyonaka, Osaka 560-0043, JAPAN}}

\newcommand{\SNU}{\affiliation{ Department of Physics, Seoul National
    University, Seoul 151-742, KOREA}}

\newcommand{\Kobe}{\affiliation{ Kobe University, Kobe, Hyogo
    657-8501, JAPAN}}

\newcommand{\UW}{\affiliation{ Department of Physics, University of
    Washington, Seattle, WA 98195-1560, USA }}

\newcommand{\UCI}{\affiliation{ Department of Physics and Astronomy,
    University of California, Irvine, Irvine, CA 92697-4575, USA }}

\newcommand{\CNU}{\affiliation{ Department of Physics, Chonnam
    National University, Kwangju 500-757, KOREA}}

\newcommand{\ICRR}{\affiliation{ Institute for Cosmic Ray Research,
    University of Tokyo, Kashiwa, Chiba 277-8582, JAPAN}}

\newcommand{\Kyoto}{\affiliation{ Department of Physics, Kyoto
    University, Kyoto 606-8502, JAPAN}}

\newcommand{\Tohoku}{\affiliation{ Research Center for Neutrino
    Science, Tohoku University, Sendai, Miyagi 980-8578, JAPAN}}

\newcommand{\KEK}{\affiliation{ High Energy Accelerator Research Organization (KEK), Tsukuba, Ibaraki 305-0801, JAPAN }}

\newcommand{\SUNY}{\affiliation{ Department of Physics and Astronomy,
    State University of New York, Stony Brook, NY 11794-3800, USA}}

\newcommand{\Okayama}{\affiliation{ Department of Physics, Okayama
    University, Okayama, Okayama 700-8530, JAPAN}}

\newcommand{\BU}{\affiliation{ Department of Physics, Boston
    University, Boston, MA 02215, USA}}

\newcommand{\Hawaii}{\affiliation{ Department of Physics and
    Astronomy, University of Hawaii, Honolulu, HI 96822, USA}}

\newcommand{\Warsaw}{\affiliation{ Institute of Experimental Physics,
    Warsaw University, 00-681 Warsaw, POLAND }}

\newcommand{\SINS}{\affiliation{ A. Soltan Institute for Nuclear
    Studies, 00-681 Warsaw, POLAND}}

\newcommand{\KU}{\affiliation{ Department of Physics, Korea
    University, Seoul 136-701, KOREA}}

\newcommand{\Niigata}{\affiliation{ Department of Physics, Niigata
    University, Niigata, Niigata 950-2181, JAPAN}}

\newcommand{\Dongshin}{\affiliation{ Department of Physics, Dongshin
    University, Naju 520-714, KOREA}}

\newcommand{\MIT}{\affiliation{ Department of Physics, Massachusetts
    Institute of Technology, Cambridge, MA 02139, USA}}

\newcommand{\TSU}{\affiliation{ Department of Physics, Tokyo
    University of Science, Noda, Chiba 278-0022, JAPAN}}

\newcommand{\Miyakyo}{\affiliation{ Department of
    Physics, Miyagi University of Education, Sendai 980-0845, JAPAN}}

\newcommand{\Duke}{\affiliation{ Department of
    Physics, Duke University, Durham, NC 27708, USA}}

% The alternative affilations will show up in the references

\newcommand{\now}{\altaffiliation{For current affiliations see\\
\texttt{http://neutino.kek.jp/present-addresses0408.ps} .}}

% not used for now
\newcommand{\ICEPP}{\altaffiliation{ International Center for Elementary Particle Physics,
    University of Tokyo, Tokyo, 113-0033, JAPAN}}

\newcommand{\KRISS}{\altaffiliation{ Korea Research Institute of Standard and Science, Yuseong, Daejeon, 305-600, KOREA}}

\newcommand{\Tokai}{\altaffiliation{ Department of Physics, Tokai
    University, Hiratsuka, Kanagawa 259-1292, JAPAN}}

\newcommand{\UPnow}{\altaffiliation{ Present address: University of
  Pittsburgh, Pittsburgh, PA 15260, USA}}

\newcommand{\Hillnow}{\altaffiliation{ Present address: California State
    University, Dominghez Hills, USA}}

\newcommand{\Jangnow}{\altaffiliation{ Present address: Seokang College,
    Kwangju, 500-742, KOREA}}

\newcommand{\Kainow}{\altaffiliation{ Present address: Department of
    Physics, University of Utah, Salt Lake City, UT 84112, USA}}

\newcommand{\Marunow}{\altaffiliation{ Present address: The Enrico Fermi
    Institute, University of Chicago, Chicago, IL 60637, USA}}

\newcommand{\Maugernow}{\altaffiliation{ Present address: California
    Institute of Technology, California 91125, USA}}

\newcommand{\Nagoyanow}{\altaffiliation{ Present address: Department of
    Physics, Nagoya University, Nagoya, Aichi 464-8602, JAPAN}}

\newcommand{\DankaSup}{\altaffiliation{ Supported by the Polish Committee
    for Scientific Research}}

\author{S.Nakayama}\ICRR
\author{C.Mauger}\now\SUNY
%\author{C.Mauger}\SUNY
\author{M.H.Ahn}\SNU
\author{S.Aoki}\Kobe
\author{Y.Ashie}\ICRR
\author{H.Bhang}\SNU
\author{S.Boyd}\now\UW
\author{D.Casper}\UCI
\author{J.H.Choi}\CNU
\author{S.Fukuda}\ICRR
% ...

\collaboration{The K2K Collaboration}\noaffiliation

%%% Abstract %%%
\begin{abstract}
Neutral current single $\pizero$ production
induced by neutrinos with a mean energy of 1.3\,GeV
is measured at a 1000\,ton water Cherenkov detector
as a near detector of the K2K long baseline neutrino experiment.
The cross section for this process
relative to the total charged current cross section
is measured to be $0.064 \pm 0.001\,(stat.) \pm 0.007\,(sys.)$.
The momentum distribution of produced $\pizero$s
is measured and is found to be in good agreement with
an expectation from the present knowledge of the neutrino cross sections.
\end{abstract}
\maketitle

%%% Introduction %%%
After the discovery of atmospheric neutrino oscillations
by Super-Kamiokande in 1998~\cite{Fukuda:1998mi},
the primal aim of current and future long baseline (LBL) experiments
using an accelerator-based neutrino beam
is more accurate determination of oscillation parameters.
The uncertainties on the knowledge of
the neutrino-nucleus cross sections
and subsequent nuclear effects in the GeV neutrino energy region
could become a severe limitation in future oscillation studies.
For reducing these systematic uncertainties,
near detectors of LBL experiments can provide
neutrino interaction data with much higher statistics
and better quality than existing data.

A single $\pizero$ event is a good signature of
neutral current (NC) neutrino interactions in the GeV region.
Especially in a water Cherenkov detector,
a decay of the $\pizero$ can be
clearly identified as two electromagnetic-showering Cherenkov rings.
The single $\pizero$ production rate by atmospheric neutrinos
could be usable to distinguish
between the $\numu \leftrightarrow \nutau$ and
$\numu \leftrightarrow \nus$ oscillation hypotheses.
The NC rate is attenuated in the case of
transitions of $\numu$'s into sterile neutrinos,
while it does not change in
the $\numu \leftrightarrow \nutau$ scenario.
An understanding of single $\pizero$ production
is also very important for
a search for electron neutrino appearance in LBL experiments
with a water Cherenkov detector,
because the most serious background to single-ring $\nue$ signals
is a single $\pizero$ event with only one ring reconstructed
due to highly asymmetric energies or
small opening angle of two $\gamma$-rays
in the $\pizero$ decay~\cite{Ahn:2004te}.

These single $\pizero$s in the GeV neutrino energy region
are mainly produced via the $\Delta$ resonance
as $\nu + N \to \nu + \Delta,\ \Delta \to N' + \pizero$,
where $N$ and $N'$ are nucleons.
Because of nuclear effects such as
Fermi motion, Pauli blocking, nuclear potential
and final state interactions,
$\Delta$ production and its decay could be different from
a simple picture of neutrino-nucleon interactions.
In addition, final state interactions of nucleons and mesons
during a traversal of nuclear matter
could largely modify the number, momenta, directions and charge states
of produced particles.
Though there exist several theoretical approaches
for modeling these processes,
their uncertainties are still large.
There exist very little experimental data for
NC single $\pizero$ production
and no reports on measurements with a water target,
which is the target matter of a far detector
in some of the LBL experiments~\cite{Ahn:2001cq,Itow:2001ee}.
It is clear that a good knowledge of
NC single $\pizero$ production cross section
and $\pizero$ momentum distribution
is required for the above studies.
In this letter, we present the first high statistics measurement of
`` NC1$\pizero$ '' interactions in water
defined as a neutral current neutrino interaction
where just a single $\pizero$ and no other mesons
are emitted in the final state from a nucleus.

%---------------------------------------------------------------
\

\begin{figure}[t]
 \includegraphics[width=3.0in]{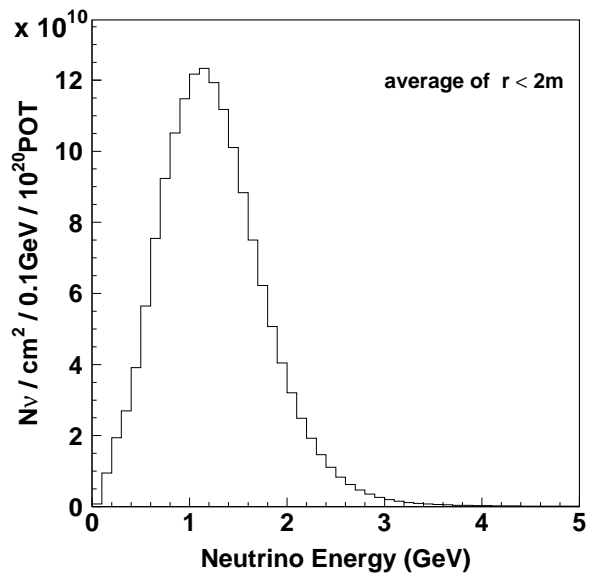}
 \caption{The energy spectrum of
 the K2K neutrino beam at 300\,m downstream from the target
 in the near site
 with a $10^{20}$ protons on target exposure
 predicted by a neutrino beam simulation.
 The spectrum is averaged within 2\,m from the beam center.}
 \label{fig:spectrum}
\end{figure}

%%% K2K Beam %%%
The KEK to Kamioka long baseline neutrino oscillation experiment (K2K)
uses an almost pure muon neutrino beam
(98.2\,\% $\numu$, 1.3\,\% $\nue$ and 0.5\,\% $\numubar$)
with a mean neutrino energy of 1.3\,GeV.
This intense wide-band beam is
produced by the KEK proton synchrotron (KEK-PS).
Protons with 12\,GeV kinetic energy extracted from the KEK-PS
are bent toward the far detector, Super-Kamiokande,
located 250\,km away from KEK
and interact with an aluminum target.
Positively charged secondary particles, mainly $\pi^+$'s,
are focused with a pair of magnetic horns
and then decay to produce a neutrino beam.
Figure~\ref{fig:spectrum} shows the energy spectrum of
the K2K neutrino beam at 300\,m downstream from the target
in the near site
with a $10^{20}$ protons on target (POT) exposure
predicted by a beam simulation~\cite{Ahn:2001cq}.
The beam simulation is validated by
a pion monitor (PIMON),
which measures the momentum and divergence of pions
just behind the second horn~\cite{Ahn:2001cq}.
The flux shape is also measured through neutrino interactions
by K2K near detectors~\cite{Ahn:2002up}.
The absolute flux estimation, however, is not easy
due to uncertainties of the primary proton beam
absolute intensity, the proton beam profile
and the proton targeting efficiency.
Instead of deriving the absolute cross section for
NC1$\pizero$ interactions,
we measure the relative NC1$\pizero$ cross section
to the total charged current (CC) cross section,
since the CC is a good experimental signature.

%%% 1kt general %%%
As one of the near detectors for K2K,
a 1,000\,ton water Cherenkov detector (1kt)
is located about 300\,m downstream of the pion production target.
The 1kt detector is a miniature of Super-Kamiokande
using the same interaction target matter
and instrumentation.
The inner volume of the 1kt detector is
a cylinder with 8.6\,m diameter and a height of 8.6\,m.
This volume is viewed by 680 photomultiplier tubes (PMTs)
of 50\,cm diameter, facing inward to detect
Cherenkov light from neutrino events.
The PMTs and their arrangement are identical to those of
Super-Kamiokande, giving a 40\,\% photocathode coverage.
The primal role of the 1kt detector is to measure
the $\numu$ interaction rate and the $\numu$ energy spectrum.
The 1kt detector also provides a good measurement of
high statistics neutrino-water interactions.
The analysis in this letter is based on the 1kt data
taken between January 2000 and July 2001,
corresponding to $3.2\times10^{19}$ POT.

The data acquisition (DAQ) system of the 1kt detector
is also similar to that of Super-Kamiokande.
A signal from each PMT is processed by
custom electronics modules called ATMs,
which are developed for the Super-Kamiokande experiment
and are used for recording
digitized charge and timing informations of each PMT hit
over a threshold of about $1/4$ photoelectrons~\cite{Fukuda:2002uc}.
The DAQ trigger threshold is about 40 hits of PMTs
within a 200\,nsec window in a 1.2\,$\mu$sec beam spill gate,
which is roughly equivalent to a signal of a 6\,MeV electron.
The analog sum of all 680 PMTs' signals (PMTSUM)
is also recorded for every beam spill by a 500\,MHz FADC
to identify multi interactions in a spill gate.
We determine the number of interactions in each spill
by counting the peaks in PMTSUM greater than
a threshold equivalent to a 100\,MeV electron signal.

Physical parameters of an event in the 1kt detector
such as the vertex position, the number of Cherenkov rings,
particle types and momenta are determined by the same algorithms as
used in Super-Kamiokande~\cite{Fukuda:1998tw,Shiozawa:1999sd}.
First, the vertex position of an event is determined by
timing information of PMT hits.
With knowledge of the vertex position,
the number of Cherenkov rings and their directions
are determined by a maximum-likelihood procedure.
Each ring is then classified as
$e$-like representing a showering particle ($e^{\pm}$,\,$\gamma$) or
$\mu$-like representing a non-showering particle ($\mu^{\pm}$,\,$\pi^{\pm}$)
using its ring pattern.
On the basis of this particle type information,
the vertex position of a single-ring event is further refined.
The momentum corresponding to each ring is determined from the intensity
of Cherenkov light.

%---------------------------------------------------------------
\

%%% FC 2ring pi0 %%%
In this analysis, neutrino interaction candidates
are selected by the following requirements :
(i) An event is triggered within a 1.2\,$\mu$sec beam spill gate.
(ii) There is no detector activity within a 1.2\,$\mu$sec window
preceding a beam spill.
(iii) Only a single event is observed by the PMTSUM peak search
in that spill.
(iv) The reconstructed vertex position is in the 25\,ton fiducial volume
defined as a 4\,m diameter, 2\,m long cylinder along the beam axis.
(v) The visible energy is larger than 30\,MeV.
A total of 60,545 events remain after cuts (i) to (v).

\begin{figure}[t]
 \includegraphics[width=3.2in]{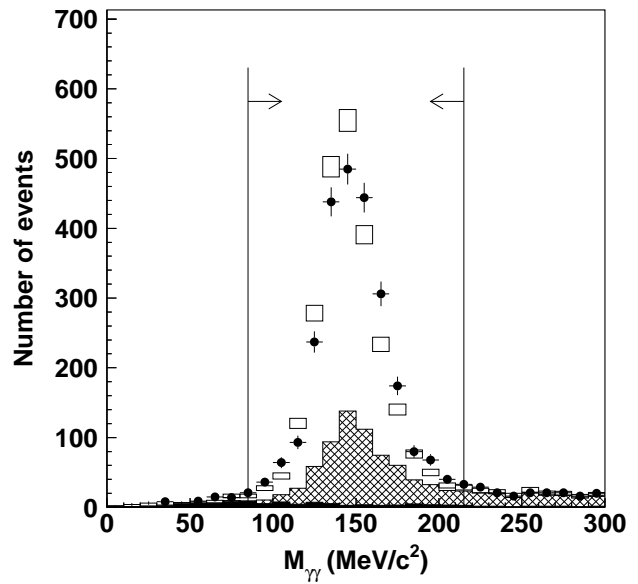}
 \caption{The invariant mass of two $e$-like ring events
 for the experimental data (black dots)
 and the neutrino Monte Carlo simulation (box histogram).
 The Monte Carlo histogram is normalized to the area of the data histogram.
 The error bars are statistical only.
 The black portion and the hatched portion in the Monte Carlo histogram
 show the non-$\pizero$ component
 and non-NC1$\pizero$ component, respectively.
 A pair of arrows shows the invariant mass cut (3) (see text).}
 \label{fig:pi0mass}
\end{figure}

Single $\pizero$ events are extracted from
the sample of fully contained (FC) neutrino events,
which deposit all of their Cherenkov light inside the inner detector.
An exiting particle deposits a large amount of energy
at the PMT of the exiting position.
Therefore, the FC events are selected by requiring
the maximum number of photoelectrons on a single PMT
at the exit direction of the most energetic particle to be less than 200.
The events with the maximum number of photoelectrons greater than 200
are identified as a partially contained (PC) event.
Because $\pizero$s mostly decay into two $\gamma$-rays,
the following criteria are further applied to the FC sample
in order to select single $\pizero$ events :
(1) The number of reconstructed rings is two.
(2) Both rings have a showering type ($e$-like) particle identification.
(3) The invariant mass is in the range of $85-215$\,MeV/$c^2$.
Figure~\ref{fig:pi0mass} shows the invariant mass distribution
of the events which satisfy cuts (1) and (2).
A $\pizero$ mass peak is clearly visible.
The peaks for both the observed data and the neutrino Monte Carlo events
are slightly shifted towards higher values compared to
the nominal value of the $\pizero$ mass, 135\,MeV/$c^2$.
This shift is due to a vertex reconstruction bias of a few dozen cm and
an energy deposit by de-excitation $\gamma$-rays from an oxygen nucleus.
The difference in peak position
between the data distribution and the Monte Carlo prediction
is within our quoted $^{+2}_{-3}$\,\%
systematic uncertainty for the absolute energy scale.
A total of 2,496 single $\pizero$ events are collected
by these criteria as shown in Table~\ref{table:selection}.
\begin{table}[t]
 \begin{center}
  \begin{tabular}{lcc}
   \hline
   \hline
                       &  Data & ~~~NC1$\pizero$ efficiency~ \\
   \hline
   ~FC                  & 45317 & 97\,\% \\
   ~Two rings           & 11117 & 57\,\% \\
   ~Both $e$-like       &  3150 & 48\,\% \\
   ~Invariant mass~~~~~ &  2496 & 47\,\% \\
   \hline
   \hline
  \end{tabular}
  \caption{The number of events after each selection
  to make the single $\pizero$ sample in 1kt data.
  The Monte Carlo efficiencies are calculated for NC1$\pizero$ interactions
  whose real vertex is in the fiducial volume.}
  \label{table:selection}
 \end{center}
\end{table}

\begin{figure}[t]
 \includegraphics[width=3.2in]{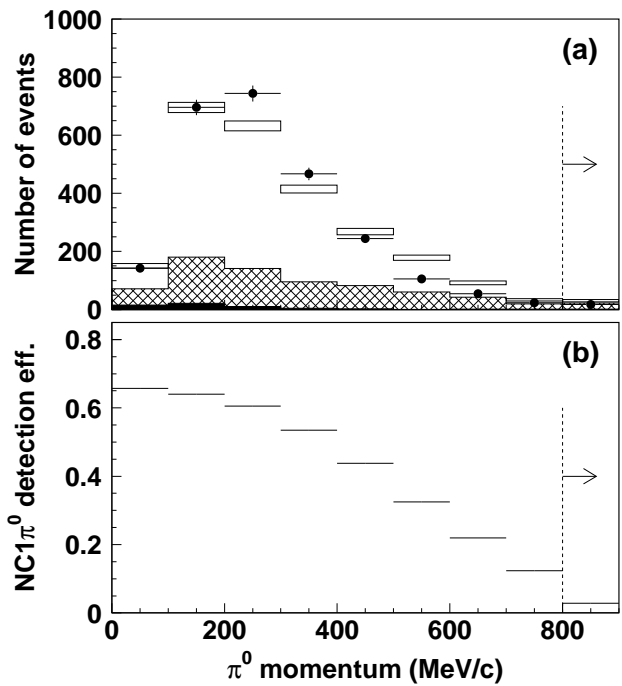}
 \caption{(a) The reconstructed $\pizero$ momentum distribution
 for the single $\pizero$ sample,
 comparing 1kt data (black dots)
 and the neutrino Monte Carlo simulation (box histogram).
 The Monte Carlo histogram is normalized to the area of the data histogram.
 The error bars are statistical only.
 The black portion and the hatched portion in the Monte Carlo histogram
 show the non-$\pizero$ component
 and non-NC1$\pizero$ component, respectively.
 (b) The detection efficiency for NC1$\pizero$ interactions
 as a function of real $\pizero$ momentum.
 The highest momentum bin in each figure integrates
 the events above 800\,MeV/$c$.}
 \label{fig:pi0momeff}
\end{figure}
Figure~\ref{fig:pi0momeff}\,-(a) shows
the reconstructed $\pizero$ momentum distribution
for the single $\pizero$ sample.
The momentum resolution for 200\,MeV/$c$ $\pizero$s
in NC1$\pizero$ events is estimated to be about 15\,MeV/$c$.
The single $\pizero$ sample contains
a background of non-NC1$\pizero$ events shown as a hatched histogram
in Figures~\ref{fig:pi0mass} and \ref{fig:pi0momeff}\,-(a).
The background contamination is estimated
by the neutrino Monte Carlo simulation
and is subtracted.

The neutrino interactions with water
are simulated by the NEUT program library.
A detailed description of NEUT can be found
in Ref.~\cite{Hayato:2002sd}.
Quasi-elastic scattering is simulated
based on the Llewellyn Smith's model~\cite{LlewellynSmith:1972zm}.
The Rein and Sehgal's model~\cite{Rein:1981wg} is used to simulate
single meson production mediated by a baryon resonance.
The decay kinematics of the $\Delta(1232)$ resonance
is also determined by the Rein and Sehgal's method.
For the decays of the other resonances,
the generated mesons are assumed to be emitted isotropically
in the resonance rest frame.
A twenty percent suppression of pion production is 
adopted for simulating pion-less $\Delta$ decay,
where the event contains only a lepton and a nucleon
in the final state~\cite{Singh:1998ha}.
The axial vector mass, $M_A$, is set to 1.1\,GeV/$c^2$
for both the quasi-elastic scattering and
the single meson production models.
Coherent pion production is simulated
using the Rein and Sehgal's model~\cite{Rein:1983pf}
with modified cross section according to
a description by Marteau et al.~\cite{Marteau:1999jp}.
For deep inelastic scattering,
GRV94 nucleon structure function~\cite{Gluck:1995uf}
with a correction by Bodek and Yang~\cite{Bodek:2002vp}
is used to calculate the cross section.
In order to generate the final state hadrons,
PYTHIA/JETSET~\cite{Sjostrand:1994yb} package is used
for the hadronic invariant mass, $W$, larger than 2.0\,GeV/$c^2$.
A custom-made program~\cite{Nakahata:1986zp} is used
for $W$ in the range of $1.3-2.0$\,GeV/$c^2$.
Nuclear effects for $\nu - ^{16}$O scattering
are also taken into account.
Fermi motion of nucleons is simulated using
the relativistic Fermi gas model.
Pauli blocking effect is considered in the simulation of
quasi-elastic scattering and single meson production.
Pions generated in $^{16}$O are tracked taking into account
their inelastic scattering, charge exchange and absorption.
The pion generation point in the nucleus is set according to
Woods and Saxon's nucleon density distribution~\cite{Woods:1954}.
The mean free path of each pion interaction is
calculated using the model by Salcedo et al.~\cite{Salcedo:1988md}.
The direction and momentum of pions after
inelastic scattering or charge exchange
are determined based on the results of a phase shift analysis
obtained from $\pi-N$ scattering experiments~\cite{Rowe:1978fb}.
Emission of de-excitation low energy $\gamma$-rays
from a hole state of residual nuclei
is also taken into account~\cite{Ejiri:1993rh}.
Outside the nucleus, particles are tracked
using GEANT and CALOR packages
except pions with momenta smaller than 500\,MeV/$c$,
which are tracked with a custom-made program~\cite{Nakahata:1986zp}
in order to consider the $\Delta$ resonance correctly.

%---------------------------------------------------------------
\

%%% NC1pi0 estimation %%%
The NC1$\pizero$ fraction in the single $\pizero$ sample,
predicted by the neutrino Monte Carlo simulation,
is estimated to be 71\,\%
(52\,\% from single $\pizero$ production via a resonance decay,
3\,\% from single $\pi^{\pm}$ production via a resonance decay
and subsequent charge exchange into $\pizero$ in the target nucleus,
10\,\% from coherent $\pizero$ production
and 4\,\% from neutrino deep inelastic scattering
where the rest of mesons are absorbed inside the nucleus).
The non-NC1$\pizero$ background,
which accounts for 29\,\% of the single $\pizero$ sample
as shown in Figures~\ref{fig:pi0mass} and \ref{fig:pi0momeff},
contains
NC $\pizero$ production where
outgoing mesons except a single $\pizero$ have low momenta (7\,\%),
CC $\pizero$ production where accompanying muon and mesons
have low momenta (9\,\%),
$\pizero$ production by a recoil nucleon or mesons
outside the target nucleus (10\,\%),
and
non-$\pizero$ background due to mis-reconstruction (3\,\%).
\begin{table*}[t]
 \begin{center}
  \begin{tabular}{lc}
   \hline
   \hline
   ~~~~~~~~~~~~~~~~~~~~~~~~~~~~~~~~~Sources & Errors (\%) \\
   \hline
   {\bf(A) Systematic uncertainties in background subtraction}       &     \\
~~~~~$M_A$ in quasi-elastic and single meson ($\pm10\,\%$)           & 0.2 \\
~~~~~Quasi-elastic scattering (total cross section, $\pm10\,\%$)     & 0.0 \\
~~~~~Single meson production (total cross section, $\pm10\,\%$)      & 0.9 \\
~~~~~Coherent pion production (model dependence)                     & 1.6 \\
~~~~~Deep inelastic scattering (model dependence)                    & 5.1 \\
~~~~~Deep inelastic scattering (total cross section, $\pm5\,\%$)     & 0.5 \\
~~~~~NC/CC ratio ($\pm20\,\%$)                                       & 3.2 \\
~~~~~Nuclear effects for pions in $^{16}$O
   (absorption, $\pm30\,\%$)                                         & 1.5 \\
~~~~~Nuclear effects for pions in $^{16}$O
   (inelastic scattering, $\pm30\,\%$)                               & 0.7 \\
~~~~~Pion production outside the target nucleus
   (total cross section, $\pm20\,\%$)                                & 2.3 \\
   {\bf(B) Systematic uncertainties in fiducial volume correction}   &     \\
~~~~~Fiducial cut                                                    & 1.6 \\
   {\bf(C) Systematic uncertainties in efficiency correction}        &     \\
~~~~~Ring counting                                                   & 5.4 \\
~~~~~Particle identification                                         & 4.2 \\
~~~~~Energy scale                                                    & 0.3 \\
   \hline
   \hline
  \end{tabular}
  \caption{Summary of the systematic errors
  on the measurement of the number of NC1$\pizero$ interactions.}
  \label{table:syserror}
 \end{center}
\end{table*}

% In practice,
% In the actual procedure,
The NC1$\pizero$ fraction is estimated
for each of the ten $\pizero$ momentum bins
shown in Figure~\ref{fig:pi0momeff}\,-(a).
The number of observed single $\pizero$ events
in each $\pizero$ momentum bin
is multiplied by the NC1$\pizero$ fraction for the corresponding momentum bin.
The systematic errors attributed to this background subtraction procedure
are estimated by assuming uncertainties in neutrino interaction models
as listed in Table~\ref{table:syserror}\,-(A).
The $M_A$ value in the models
for quasi-elastic scattering and single meson production
is varied by $\pm 10\,\%$
from a central value of 1.1\,GeV/$c^2$.
The error due to the uncertainty on
coherent pion production cross section is estimated by
removing the cross section modification by Marteau et al..
For deep inelastic scattering, the effect of
the correction on nucleon structure function
according to Bodek and Yang's description is evaluated.
The systematic errors due to nuclear effect uncertainties
are evaluated by varying the probabilities of
pion absorption and inelastic scattering (including charge exchange)
in $^{16}$O nucleus independently by $\pm 30\,\%$.

After estimating the number of NC1$\pizero$ interactions
in each $\pizero$ momentum bin,
we apply a fiducial volume correction.
The number of NC1$\pizero$ interactions
in the Monte Carlo single $\pizero$ sample
increases by about 2\,\%
if real vertices are used instead of reconstructed vertices
in the fiducial volume selection.
The data are then multiplied by 1.02 in order to derive
the number of NC1$\pizero$ interactions in the true 25\,ton fiducial volume
before the efficiency correction is applied.
The systematic error in the fiducial volume correction
is about 2\,\% as shown in Table~\ref{table:syserror}\,-(B),
which is estimated from the difference
in the reconstructed vertex distribution of single $\pizero$ events
for 1kt data and the neutrino Monte Carlo simulation.

Finally, the number of NC1$\pizero$ interactions
in the true 25\,ton fiducial volume is corrected for
the detection efficiency in a bin by bin manner.
Figure~\ref{fig:pi0momeff}\,-(b) shows
the detection efficiency for NC1$\pizero$ interactions
as a function of real $\pizero$ momentum.
The inefficiency for higher momentum $\pizero$s
is primarily due to reconstruction of only one ring
for the $\pizero$ decay with highly asymmetric energies
or small opening angle of two $\gamma$-rays.
The overall detection efficiency for NC1$\pizero$ interactions,
estimated by the neutrino Monte Carlo simulation,
is 47\,\% as shown in Table~\ref{table:selection}.
The systematic errors from the efficiency correction
are due to uncertainties of reconstruction algorithms
such as ring counting, particle identification
and energy scale as listed in Table~\ref{table:syserror}\,-(C).

\begin{figure}[h]
 \includegraphics[width=3in]{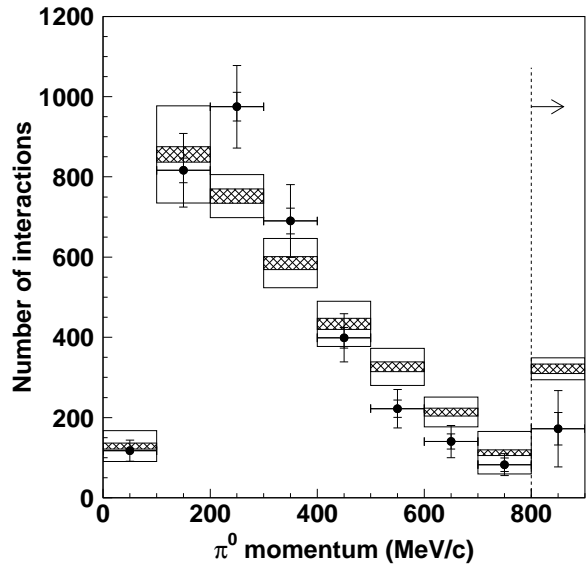}
 \caption{The momentum distribution
 of NC1$\pizero$ events
 in the 25\,ton fiducial volume (black dots).
 The inner and outer error bars attached to data points
 show statistical errors and total errors
 including systematic errors, respectively.
 The distribution predicted by the neutrino Monte Carlo simulation
 is also shown as a box histogram for comparison.
 The size of inner boxes represents Monte Carlo statistical errors.
 The size of outer boxes represents
 the uncertainty of the distribution shape due to
 neutrino interaction model ambiguity.
}
 \label{fig:pi0mom-corr}
\end{figure}
By a series of corrections for the background subtraction,
the true/reconstruction fiducial difference
and the detection efficiency,
the true number of NC1$\pizero$ interactions in our data
is measured to be $(3.61 \pm 0.07 \pm 0.36)\times 10^3$
in the 25\,ton fiducial volume.
Figure~\ref{fig:pi0mom-corr} shows
the measured momentum distribution of NC1$\pizero$
after all corrections,
compared with the distribution predicted by
the neutrino Monte Carlo simulation.
The Monte Carlo histogram is normalized
by the number of total neutrino events in the fiducial volume,
which are selected by cuts (i) to (v) as previously described.
The size of outer boxes for the Monte Carlo histogram represents
the uncertainty of the distribution shape of NC1$\pizero$ momentum
due to neutrino interaction model ambiguity,
where the largest source is nuclear effects for pions in $^{16}$O.
The measured distribution is in reasonably good agreement
on the Monte Carlo estimation.

%---------------------------------------------------------------
\

%%% Denominator %%%
As previously described,
we use CC interactions for normalization
in order to derive the relative cross section
for NC1$\pizero$ interactions.
To make a CC enriched sample,
FC $\mu$-like events and PC events are selected
from the 1kt data during the same period.
By using the CC enriched sample as a normalization,
the uncertainty of the neutrino energy spectrum~\cite{Ahn:2002up}
is almost canceled in the measurement
since the expected mean energy of neutrinos producing the CC sample,
1.45\,GeV, is almost same as that of neutrinos producing
the $\pizero$ sample, 1.50\,GeV.
The sample consists of
22,612 FC single-ring $\mu$-like events,
12,386 FC multi-ring events with the most energetic ring
identified as $\mu$-like and
15,228 PC events,
resulting in a total of 50,226 events.
The $\numu$CC fraction in this sample
is estimated to be 96\,\% by the neutrino Monte Carlo simulation
(96.5\,\% for the FC single-ring $\mu$-like sample,
91.2\,\% for the FC multi-ring $\mu$-like sample and
98.5\,\% for the PC sample).
The rest 4\,\% of the sample is mostly
composed of NC interactions accompanying an outgoing charged pion
above its Cherenkov threshold.
The fiducial volume correction factor is estimated to be 1.02.
The detection efficiency for $\numu$CC interactions
by this selection is estimated to be 85\%.
The inefficiency mainly comes from
mis-identification of the ring type in multi-ring events
and $\sim 100$\,MeV visible energy threshold
by peak counting of the PMTSUM signal.

By applying non-$\numu$CC backgound subtraction,
fiducial volume correction and detection efficiency correction,
the number of $\numu$CC neutrino interactions
during the analysis period is measured to be
$(5.65 \pm 0.03 \pm 0.26)\times 10^4$
in the 25\,ton fiducial volume.
The estimated systematic errors are
4\,\% from the uncertainty of vertex reconstruction,
1\,\% from the uncertainty of neutrino interaction models,
1\,\% from the uncertainty of particle identification and
1\,\% from the uncertainty of absolute energy scale.

By taking the ratio,
the relative cross section for NC1$\pizero$ interactions
to the total $\numu$CC cross section
is measured to be $0.064 \pm 0.001\,(stat.) \pm 0.007\,(sys.)$.
Our neutrino interaction models predict the ratio to be 0.065,
which results in good agreement.
For reference,
the total $\numu$CC cross section
is calculated to be $1.1 \times 10^{-38}$\,cm$^2$/nucleon
in the neutrino Monte Carlo simulation
by averaging over the K2K neutrino beam energy.

%---------------------------------------------------------------
\

%%% Conclusion %%%
In summary, we have measured
the rate and the $\pizero$ momentum distribution
of NC single $\pizero$ production
by neutrinos with a mean energy of 1.3\,GeV
at the K2K 1kt water Cherenkov detector.
The cross section ratio to total $\numu$CC neutrino interaction
is obtained as $0.064 \pm 0.001\,(stat.) \pm 0.007\,(sys.)$,
showing good agreement with
the prediction by our neutrino interaction models.
This measurement provides essential information
for an understanding of $\pizero$ production in water
at the 1\,GeV neutrino energy region,
which is relevant for present and future
neutrino oscillation experiments.

%%% Acknowledgment %%%
%\newpage
We thank the KEK and ICRR Directorates for their strong support and
encouragement.  K2K is made possible by the inventiveness and the
diligent efforts of the KEK-PS machine and beam channel groups.
We gratefully acknowledge the cooperation of
the Kamioka Mining and Smelting Company.
This work has been supported by the Ministry of Education,
Culture, Sports, Science and Technology, Government of Japan and its
grants for Scientific Research, the Japan Society for Promotion of
Science, the U.S. Department of Energy, the Korea Research
Foundation, the Korea Science and Engineering Foundation,
the CHEP in Korea, and Polish KBN grant 1P03B03826 and 1P03B08227.

\end{document}